\documentclass{aastex631}

\begin{document}

\title{Exploration of Halo Substructures in IoM Space with \textit{Gaia} DR3}
\shorttitle{Halo Substructures with \textit{Gaia} DR3}
\shortauthors{Liu et al.}

\correspondingauthor{Cuihua Du}
\email{ducuihua@ucas.ac.cn}

\author{Haoyang Liu}
\affiliation{College of Astronomy and Space Sciences, University of Chinese Academy of Sciences, Beijing 100049, P.R. China}

\author{Cuihua Du}
\affiliation{College of Astronomy and Space Sciences, University of Chinese Academy of Sciences, Beijing 100049, P.R. China}

\author{Dashuang Ye}
\affiliation{College of Astronomy and Space Sciences, University of Chinese Academy of Sciences, Beijing 100049, P.R. China}

\author{Jian Zhang}
\affiliation{College of Astronomy and Space Sciences, University of Chinese Academy of Sciences, Beijing 100049, P.R. China}

\author{Mingji Deng}
\affiliation{College of Astronomy and Space Sciences, University of Chinese Academy of Sciences, Beijing 100049, P.R. China}



\begin{abstract}

Using kinematic data from the Gaia Data Release 3 catalog, along with metallicity estimates robustly derived from Gaia XP spectra, we have explored the Galactic stellar halo in search of both known and potentially new substructures.
By applying the HDBSCAN clustering algorithm in IoM space (i.e. $E,L_{z}$ and $L_{\perp}$$ = \sqrt{L_{x}^2+L_{y}^2}$), we identified 5 previously known substructures: Gaia-Sausage-Enceladus (GSE), Helmi Streams, I'itoi + Sequoia and Hot Thick Disc. We additionally found NGC 3201 and NGC 5139 in this work, and NGC 3201 shares similar distributions in phase space and metallicties to Arjuna, which possibly implies that they have the same origin. Three newly discovered substructures are Prograde Substructure 1 (PG1), Prograde Substructure 2 (PG2) and the Low Energy Group. PG1, with a higher $V_{\phi}$ than typical GSE member stars, is considered as either a low eccentricity and metal-rich part of GSE or part of the metal-poor disc. PG2, sharing kinematic similarities with Aleph, is thought to be its relatively highly eccentric component or the mixture of Aleph and disc. The Low Energy Group, whose metal-poor component of metallicity distribution function has a mean value [M/H] $\sim$ $-$1.29 (compared to that of Heracles [M/H] $\sim$ $-$1.26), may have associations with Heracles.

\end{abstract}

\keywords{Galaxy stellar halos(598) --- Galaxy kinematics(602) --- Galaxy dynamics(591)}


\section{Introduction} \label{sec:intro}

In the widely accepted cosmological model ($\Lambda$-CDM model), spiral galaxies such as the Milky Way typically evolve through a hierarchical process, gradually assembling from numerous mergers with smaller galaxies \citep{White1978,Springel2005}. Simulations have predicted that the debris formed when the progenitor galaxies were tidally disrupted, are deposited in the Galactic stellar halo \citep{Yanny_2000,Majewski_2003,Zolotov_2009}. Studying the kinematic and chemical properties of these accreted substructures will provide us rich information of the formation history of the Milky Way (known as  ``Galactic archaeology") as well as information on progenitor galaxies \citep{Johnson1996,HElim1999,Bullock_2005}.

The most famous  ``structure" found in the last century is Sagittarius stream formed by the disrupting Sagittarius dwarf galaxy \citep{Ibata1994,Majewski_2003}. Other substructure fossils can be found in the phase space due to the long relaxation time, with integrals of motions (IoM) memorizing the previous kinematics. Helmi Streams \citep{HElim1999} was the first substructure found in IoM space, and later followed by Gaia-Sausage-Enceladus \citep{Belokurov2018,Helmi2018}, Thamnos \citep{koppelman_2019}, Sequoia \citep{myeong_2019}, Arjuna and I'itoi \citep{Naidu_2020}, Aleph \citep{Naidu_2020}, Pontus \citep{Malhan_2022}, LMS-1 (Wukong) \citep{Yuan_2020,Naidu_2020}, Cetus \citep{Newberg2009}, Heracles \citep{Horta2021} as well as the newly discovered metal-poor substructures Shiva and Shakti \citep{kHYATI2024}.

Most of the substructures are associated with accretion events. Gaia-Sausage-Enceladus is associated with the ancient accretion of a single  massive satellite  $\sim$10 Gyrs ago \citep{Belokurov2018,Gallart2019}, with estimated initial stellar mass of 5 $\times$ 10$^8$ - 5 $\times$ 10$^9 M_{\sun}$ \citep{Mackereth2019,Vincenzo2019}. The evidence of this major merger is the existence of Splash stars \citep{Belokrov2020}, which were heated away from the proto-disc during the merger to have more halo-like kinematics. \citet{Fattahi19} used simulations to find a single dwarf galaxy with a stellar mass of 10$^9$ - 10$^{10} M_{\sun}$ 6 - 10 Gyr ago, could contribute the metal-rich and highly eccentric halo, which is possibly associated with the Splash. Sequoia is a high energy retrograde substructure, probably accreted by the Milky Way \citep{myeong_2019}. The progenitor galaxy of Sequoia is estimated to have a stellar mass of 5 $\times$ 10$^7$ $M_{\sun}$ and a total mass of $\sim$ 10$^8$ $M_{\sun}$, and Sequoia is distinguished from GSE for having higher [Al/Fe] values. Helmi Stream probably originates from a dwarf galaxy of a total mass $\sim$ 10$^8$$M_{\sun}$ that was accreted 5-8 Gyrs ago \citep{HElim1999}. Moreover, \citet{Donlon2022} found the stellar halo is built up by more than one single radial event. Later \citet{Donlon2023} have also analyzed the components in the local halo using the Bayesian Gaussian mixture model regression algorithm to characterize the local stellar halo, and they found the data fit the model best with four components, which are Virgo Radial Merger (VRM), Cronus, Nereus, and Thamnos. They suggest that the GSE merger event is probably a combination of these four components. All these mentioned findings provide further proofs for how our Milky Way evolves through the hierarchical merger tree.

The \textit{Gaia} Mission \citep{Gaia2016Collaboration,GAIA2023} has provided accurate stellar astrometry and  photometry measurements from the whole sky. Other massive surveys also serve as supplements to each other, such as Apache Point Observatory Galactic Evolution Experiment (APOGEE; Wilson et al. \citeyear{wilson_2019}, Abdurro'uf et al. \citeyear{Abdurro2022}), Galactic
Archaeology with High Efficiency and Resolution Multi-Element
Spectrograph (GALAH; De Silva et al. \citeyear{Desilva2015}), Sloan Exploration of Galactic Understanding and Evolution (SEGUE; Yanny et al. \citeyear{2009AJYanny}), the Radial Velocity Experiment (RAVE; Steinmetz et al. \citeyear{Steinm_2006}), Large sky Area Multi-Object fiber Spectroscopic Telescope (LAMOST; Cui et al. \citeyear{Cui_2012}).

It is a common approach to study the substructures of the stellar halo using clustering algorithm. \citet{Ye2024} used Shared Nearest Neighbor (SNN) algorithm to study the very metal-poor (VMP) stars in the local halo, and found VMP stars of metal-weak Thick Disc (MWTD) might be accreted from a dwarf galaxy. \citet{shank2022} used HBDSCAN (as we did in this study) in ($E, J_{r}, J_{\phi}, J_{z}$) space to analyze the metal-poor stars from RAVE DR6, discussing the associations between dynamically tagged groups (DTGs) and substructures. There are also other authors who did the clustering algorithm in IoM space. \citet{DOdd_2022} utilized a single linkage-based clustering algorithm in IoM space to study the substructures in the solar vicinity, while \citet{Lara2022} developed a data-driven method for clustering in IoM space to further characterize the substructures with 6D phase-space information. However, the samples in the previous literature could be limited after cross-matching with data like APOGEE, GALAH to obtain reliable metallicities, which is not adopted in our work in case of the possibly missing substructures in those data.

In this study, based on the kinematic data from \textit{Gaia} DR3 \citep{Gaia2016Collaboration,GAIA2023} with precise and well-trained metallicity data from \textit{Gaia} XP spectra \citep{Andrae_2023}, we search for potential substructures in the IoM space. The paper is constructed as follows: in Section \ref{data}, we describe the quality cut from \textit{Gaia} DR3. Section \ref{method} gives a brief introduction on HDBSCAN algorithm, and Section \ref{result} shows the clustering results and corresponding substructures. Finally we discuss the results and give a comprehensive summary in Section \ref{discussion}. 

\begin{figure*}
    \includegraphics[width=\textwidth]{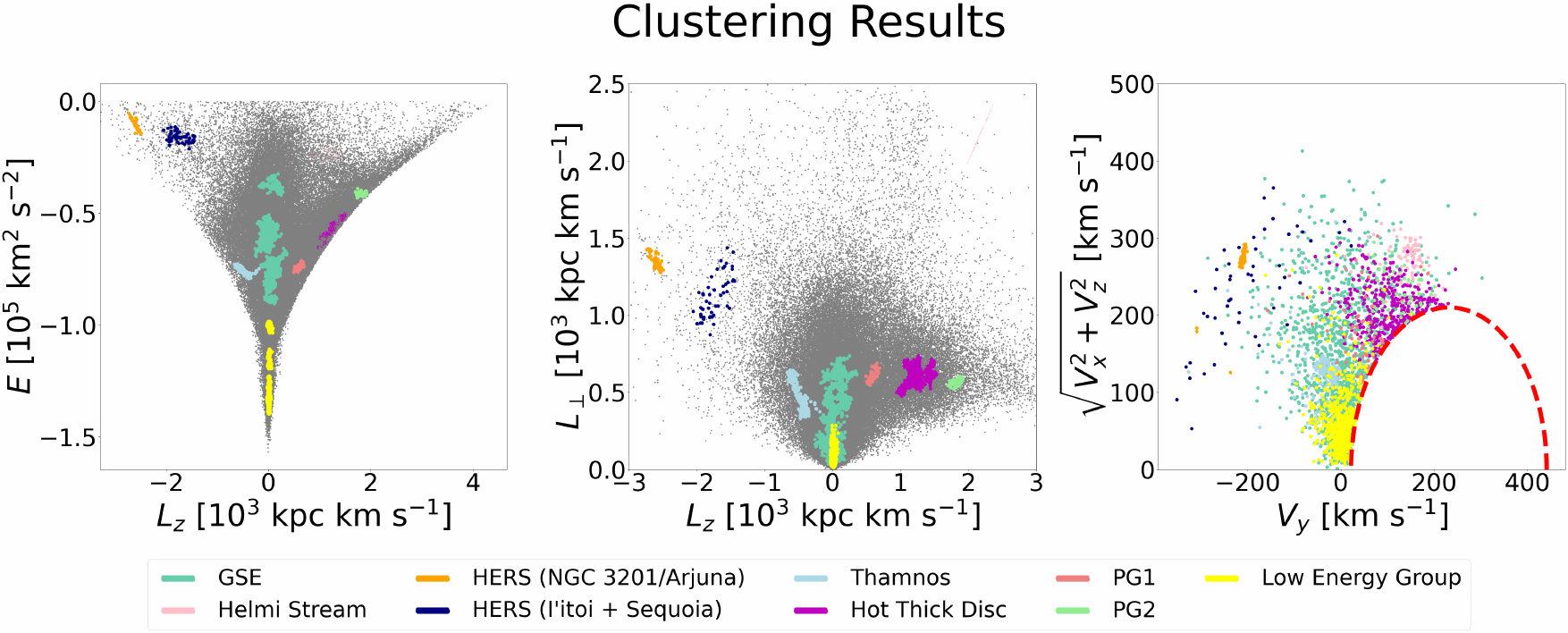} 
    \caption{30 groups generated by HDBSCAN are divided into substructures in $E-L_{z}$ plane, $L_{\perp}-L_{z}$ plane and in Toomre diagram (from left to right), including previously known  and unknown substructures: GSE (medium aquamarine), Helmi Streams (pink), Arjuna (orange), I'itoi$+$Sequoia (navy), Thamos (light blue), Hot Thick Disc (magenta), as well as three unknown debris PG1 (light coral), PG2 (light green), Low Energy Group (yellow). The red dashed line indicates $|V-V_{\text{LSR}}|$ = 210 km/s.  }
    \label{ClusterR}
\end{figure*}

\section{data} \label{data}
For this work, we use \textit{Gaia} DR3 Catalogue \citep{Gaia2016Collaboration,GAIA2023} with distances from \texttt{external.gaiaedr3\_distance} estimated by \citet{Bailer-Jones_2021} \footnote{This is available in \textit{Gaia} archive: \url{https://gea.esac.esa.int/archive/}}. For corresponding metallicities, we use the robust data-driven metallicity estimates for around 175 million stars from \textit{Gaia} XP Spectra using XGboost algorithm by \citet{Andrae_2023} \footnote{This is available in zenodo: \url{https://zenodo.org/records/7945154}}, with an excellent mean precision of 0.1 dex. After all data are matched, the quality cuts are applied as follows:
\\
 \hspace*{2em} (1)  \texttt{ruwe < 1.2}
\\
 \hspace*{2em} (2)  \texttt{radial\_velocity\_error < 10 km/s}
\\
 \hspace*{2em} (3)  \texttt{parallax\_over\_error > 5}
\\
 \hspace*{2em} (4)  \texttt{6 mag $\leq$ phot\_g\_mean\_mag $\leq$ 21 mag}
\\
 \hspace*{2em} (5)  \texttt{astrometric\_excess\_noise < 2}
\\
 \hspace*{2em} (6)  \texttt{visibility\_periods\_used > 10}
\\
 \hspace*{2em} (7)  \texttt{rv\_expected\_sig\_to\_noise > 10}
\\
 \hspace*{2em} (8)  \texttt{$\frac{\texttt{r\_hi\_geo} - \texttt{r\_lo\_geo}}{\texttt{r\_med\_geo}}$ < 0.4}
\\
 \hspace*{2em} (9)  \texttt{duplicated\_source = 0}

where (1) - (7) are thresholds that ensure spectral quality, accurate radial velocities and reliable parallaxes' zero-point (based on recommondations from \citet{Lindegren2021}. As for (8), \texttt{r\_hi\_geo} and \texttt{r\_lo\_geo} represent the 84th and 16th percentile of geometric distances respectively, while \texttt{r\_med\_geo} indicates the median values, and this threshold limits the distance error to be under 0.2. In order to avoid observational, cross-matching or processing problems, or stellar multiplicity, we also give the condition (9). This produces a whole sample of 8,537,332 stars. 
Orbital parameters including energy and angular momentum are calculated using \texttt{gaply} under the axisymmetric potential model \texttt{MWPotential2014} \citep{Bovy2015}. We assume the Local Standard of Rest (LSR) velocity $V_{\text{LSR}}$ is 232 km/s, a radius distance of the Sun is 8.21 kpc \citep{Mc17}, and the height above the mid-plane is 20.8 pc \citep{Bennet&Bovy}. The solar motion is $[U_{\sun}, V_{\sun}, W_{\sun}] = [11.1, 12.24, 7.25]$ km/s \citep{schonrich}. We integrate the orbits forward in 10 Gyr to obtain kinematic paramenters for stars, like energy, angular momentum, eccentricity and $z_{\text{max}}$ (the maximum height above the Galactic plane). We apply the right-handed Galactocentric frame of reference, where x, y and z indicate the Cartesian coordinates; R is the cylindrical radius; $\phi$ represents the azimuthal angle. For halo stars selection, we use $|V-V_{\text{LSR}}|$ $>$ 210 km/s and apply a $z_{\text{max}}$ cut at $z_{\text{max}}$ $>$ 2.5 kpc, where the latter cut is to remove possible thin disc stars in \citet{Ouxiaowei}. At last, stars with positive energy are removed and now we have 88,068 stars in total.

\section{METHOD}\label{method}
For given halo samples, we applied Hierarchical Density-Based Spatial Clustering of Applications with Noise (HDBSCAN) in three-dimensional  Integrals of Motion (IoM) space, including energy (\textit{E}), angular momentum z-component ($L_{z}$) and in-plane component of angular momentum ($L_{\perp}$$ = \sqrt{L_{x}^2+L_{y}^2}$), in which $L_{\perp}$ is not fully conserved. The reason for the choice of IoM space rather than action space is that it is more intuitive with those familiar quantities.

HDBSCAN, as an improved density-base and hierarchical clustering method developed by \citet{Campello}, can extract the flat portions from the simplified tree of significant clusters. The algorithm clusters the data according to the density of points without making any assumptions on the number of clusters and the distribution of data, thus allowing the existence of noise. HDBSCAN shows the best performance, when \citet{Brauer_2022} tried to examine the results of a suite of clustering algorithms on substructures left by dwarf galaxies in a cosmological simulation from \citet{Griffen_2016}.

Bascially, clusters could be found through the definition of the mutual reachability distance that is shown below:
\begin{equation}
    d_{\text{mreach}-k}(a,b) = \max \{ \text{core}_{k}(a), \text{core}_{k}(b), d(a,b) \}
\end{equation}
where $\text{core}_{k}(x)$ stands for the radius of the point $x$ when it has k nearest neighbors (i.e. the core distance), and $d(a,b)$ is the distance of two points $a$ and $b$ under a chosen metric. A mutual reachability distance metric is set up based on the definition, and thus a minimum spanning tree is established. This tree will be sorted by the values of mutual reachability distance, and the clusters with fewer points than the hyper-parameter \texttt{min\_cluster\_size} will be discarded and treated as noise. The left clusters will be kept if they are stable against further splitting.

\begin{figure*}
    \includegraphics[width=\textwidth]{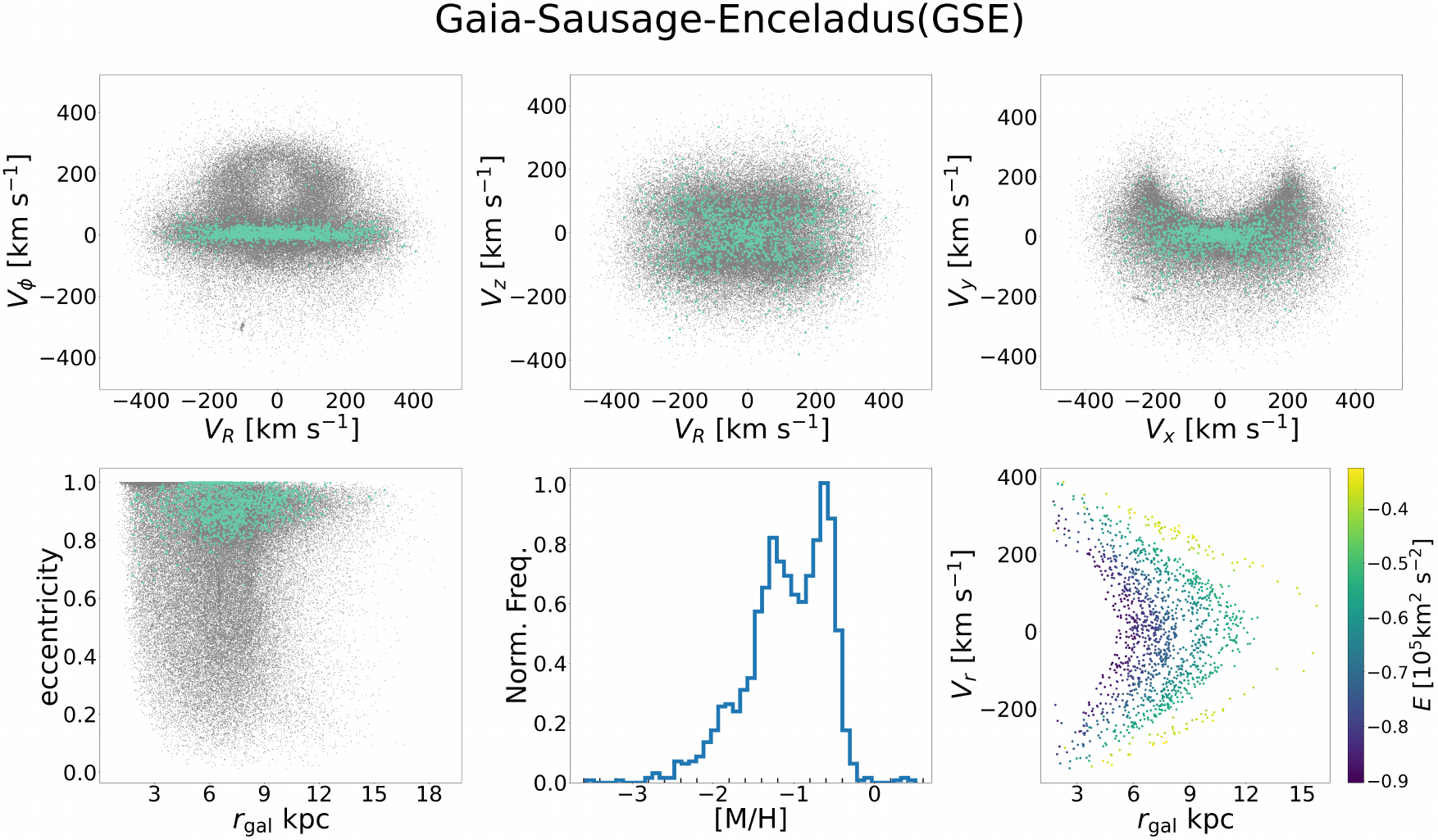} %
    \caption{Top row:the velocity distributions in $V_{R}-V_{\phi}$, $V_{R}-V_{x}$, $V_{x}-V_{y}$ space, where sausage shape is visible in $V_{R}-v_{\phi}$ space. Bottom row: The eccentricity distribution along the galacocentric distance $r_{\text{gal}}$ and metallicity distribution of GSE with two noticeable peaks that peak around $-$1.25 and $-$0.7 respectively, where the latter peak is related to eccentric Splash stars. The chevron-like shape in $V_{r}-r_{\text{gal}}$ space is shown and color-coded by energy.}
    \label{GSE}
\end{figure*}

\begin{figure*}
    \includegraphics[width=\textwidth]{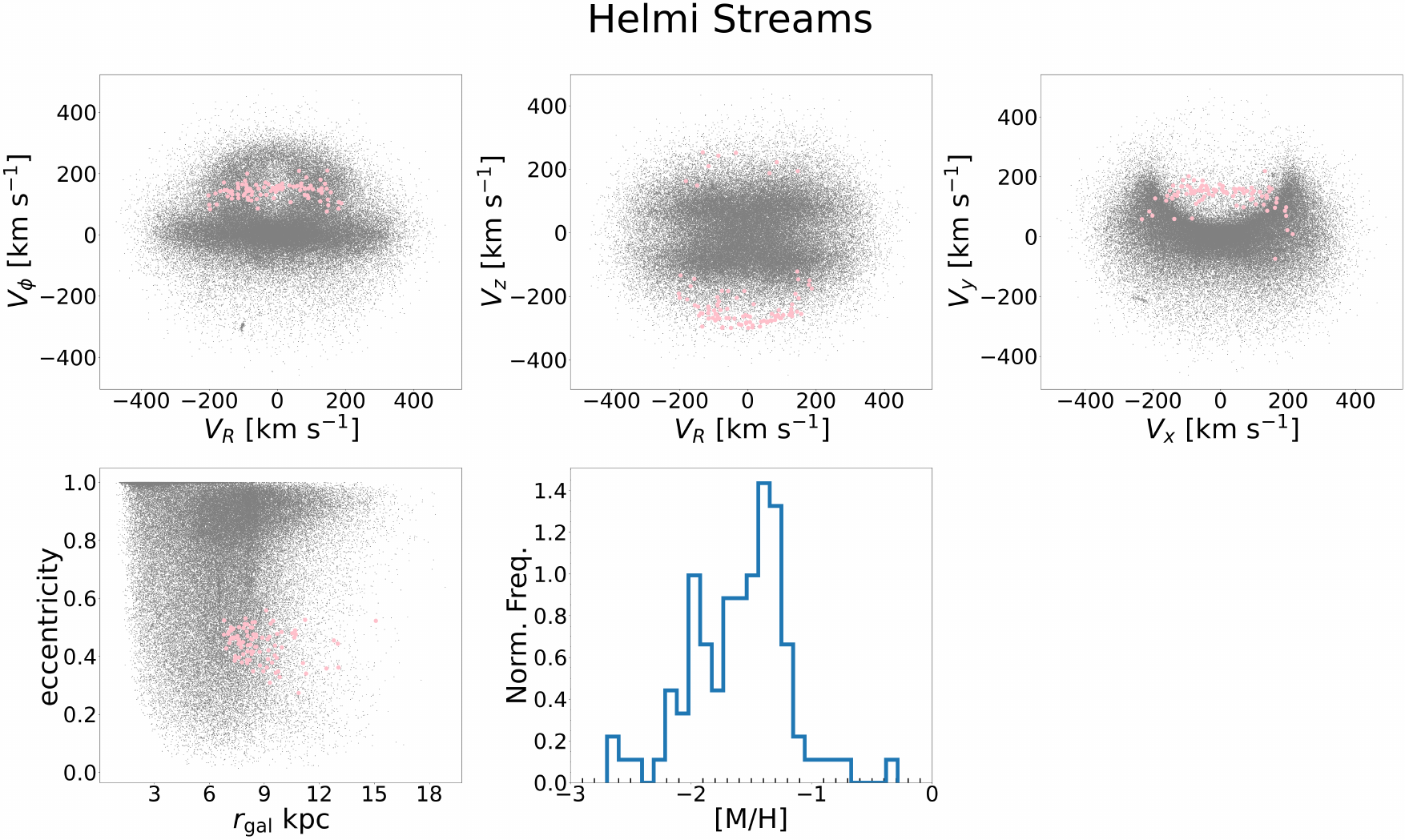} %
    \caption{Top row:the velocity distributions in $V_{R}-V_{\phi}$, $V_{R}-V_{x}$, $V_{x}-V_{y}$ space.  Bottom row: The eccentricity distribution along the galacocentric distance $r_{\text{gal}}$ and metallicity distribution of Helmi Streams. Most Helmi Streams member stars show negative $V_{z}$ velocities, as well as metal-poor characteristics of MDF.}
    \label{HS}
\end{figure*}

\begin{figure*}
    \includegraphics[width=\textwidth]{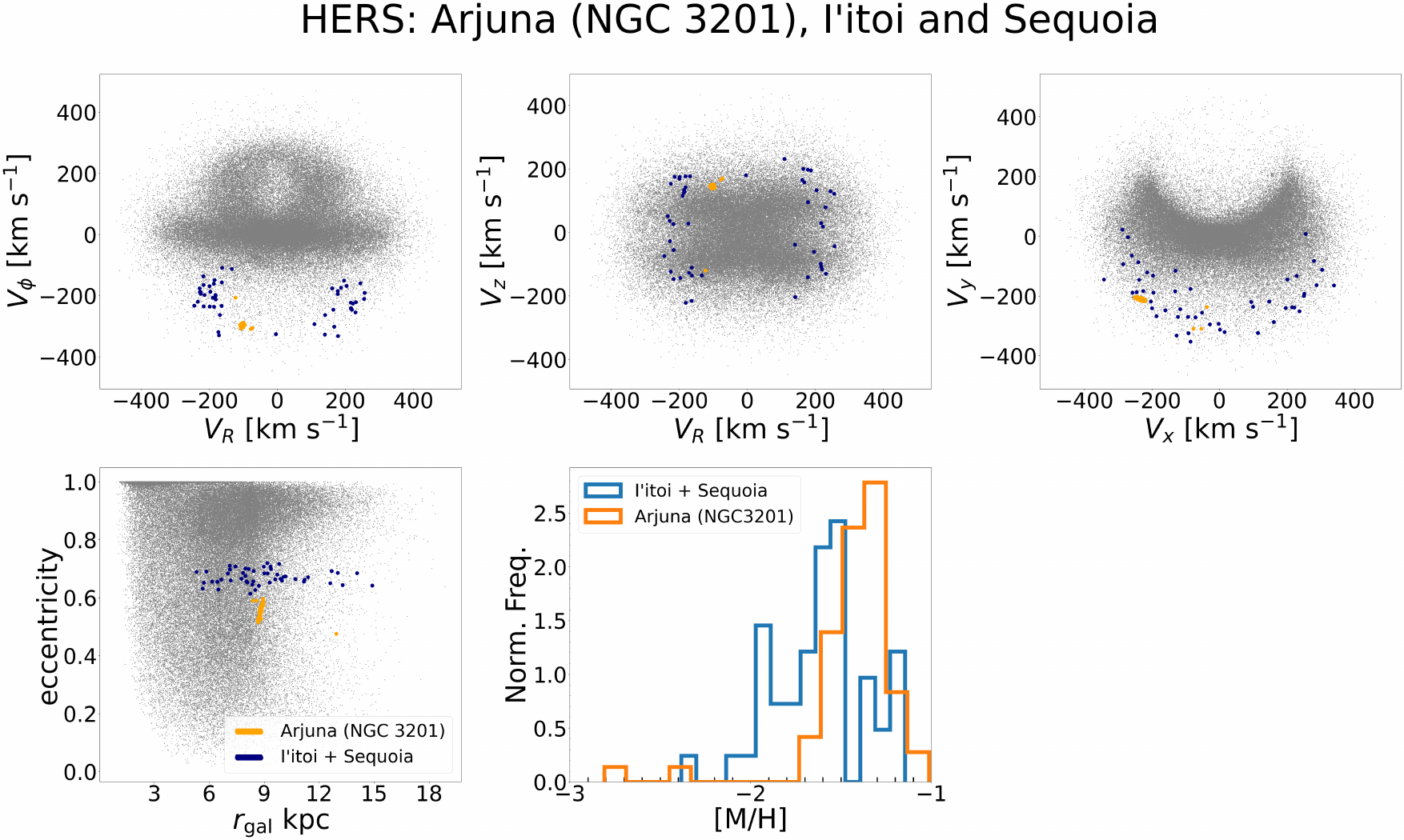} %
    \caption{Top row:the velocity distributions in $V_{R}-V_{\phi}$, $V_{R}-V_{x}$, $V_{x}-V_{y}$ space. Bottom row: The eccentricity distributions along the galacocentric distance $r_{\text{gal}}$ and metallicity distributions of HERS. The positions of all NGC 3201/Arjuna stars are nearly identical in both phase and velocity space (see Figure.~\ref{ClusterR}), suggesting that they may originate from a globular cluster. Another notable point is that I'itoi and Sequoia exhibit stable eccentricities around 0.7, with their $r_{\text{gal}}$ ranging from 6 kpc to 15 kpc. }
    \label{HERS}
\end{figure*}

In the HDBSCAN algorithm \citep{McInnes2017hdbscanHD}, we mainly focus on two hyper-parameters here: \texttt{min\_cluster\_size} and \texttt{min\_samples}. The first one decides the minimum number of a cluster and the latter one is related to the noise (more points will be included in the cluster if this parameter is set to a lower value). Considering the large amount of our samples and after testing on the two parameters, we set \texttt{min\_cluster\_size} to be 50, which is different from the normal choice of 20, and \texttt{min\_samples} is set to be 25. The method selection is set to \texttt{leaf} to achieve fine-grained and homogeneous clusters, which aids in uncovering existing and potential new substructures within the data, as well as enabling the tracing of merger tree nodes and the visualization of condensed trees. Finally after removing stars with cluster probabilities below 0.9, we end up with 30 groups containing 2,932 stars.

\section{Results}\label{result}
We have successfully identified various groups and uncovered both known substructures and potential unknown debris by analyzing their positions in the phase space, as well as their precise metallicity, which has been effectively modeled using XGBoost.  The existing found substructures are Gaia-Sausage-Enceladus (GSE), Helmi Streams, Thamnos, high enegry retrograde substructures (HERS, which includes Arjuna, I'itoi and Sequoia), hot thick disc. The three unknown debris are prograde substructure 1 and prograde substructure 2 (hereinafter referred to as PG1 and PG2, respectively), and the Low Energy Group (named by the fact of having the lowest energy in $E-L_{z}$ plane). 

The results of discovered substructures are shown in Figure.~\ref{ClusterR}, where 30 clusters have been divided into substructures based on phasic positions and metallicities. The detailed analyses on each substructure are provided in the following subsections.

\subsection{Gaia-Sausage-Enceladus}\label{G-S-E}
The Gaia-Sausage-Enceladus (GSE) is the most thoroughly researched substructure, previously found by \citet{Belokurov2018} and \citet{Helmi2018}. It has been the subject of extensive research in subsequent studies \citep{Koppelman_2018,myeong_2018,Haywood_2018,Mackereth2019}, and is estimated to constitute approximately 40 percent of the stellar halo in the vicinity of the Sun \citep{MackerBovy2020}.

What creates the name is because of the blob-shaped/sausage-shaped distribution in the $V_{r}-V_{\phi}$ plane, and they possess highly eccentric orbits, with the majority having an eccentricity greater than 0.7. Due to the above properties, \citet{Naidu_2020} simply selects GSE members that have eccentricities higher than 0.7, while \citet{Horta2023} also puts a limitation on $|L_{z}|< $ 500 kpc km s$^{-1}$. 

Here, we identify clusters as the GSE members if the 16th percentile of their eccentricity exceeds 0.7 and their angular momentum meets the criterion of $|L_{z}|< $ 500 kpc km s$^{-1}$; therefore, 21 groups have been initially identified as GSE members. However, we noticed that 6 groups have extremely eccentric orbits, with eccentricities approaching \textit{e} $\sim$ 1 (and r$_{\text{gal}} < \sim$ 5 kpc). These are assigned to the Low Energy Group due to their exceptionally low energy, which may be associated with the substructure Heracles, firstly discovered by \cite{Horta2021}. As a result, GSE members contain 15 groups with 1,358 stars.

In Fig.~\ref{ClusterR}, we can easily see the GSE distribution in the phase space where engery occupies the area ($-$ 90000 km$^2$ s$^{-2} < E < -$ 30000 km$^2$ s$^{-2}$) with $|L_{z}| <$ 500 kpc km s$^{-1}$. In Fig.~\ref{GSE}, the long and thin chevron-like shape is also re-appeared in bottom-right panel, which is due to the incomplete phase mixing in the last massive merger event \citep{Belokurov_2023}, or perhaps the stars trapped in resonances with the Galactic bar \citep{Dillamore}.
The sausage-shaped distribution is shown in top-middle panel, where GSE members show a wide span of $V_{r}$ from $\sim$ $-$300 km/s to $\sim$ 300 km/s. The metallicity distribution function (MDF) of GSE also exhibits a broad range with two noticeable peaks, at [M/H] $\sim$ $-$1.25 and at [M/H] $\sim$ $-$0.7 respectively. The second peak is probably caused by the contamination by Splash stars with [M/H] $> -$ 0.7 that were heated away from the proto-disc due to the last major merger event \citep{Belokrov2020}.

\begin{figure*}
    \includegraphics[width=\textwidth]{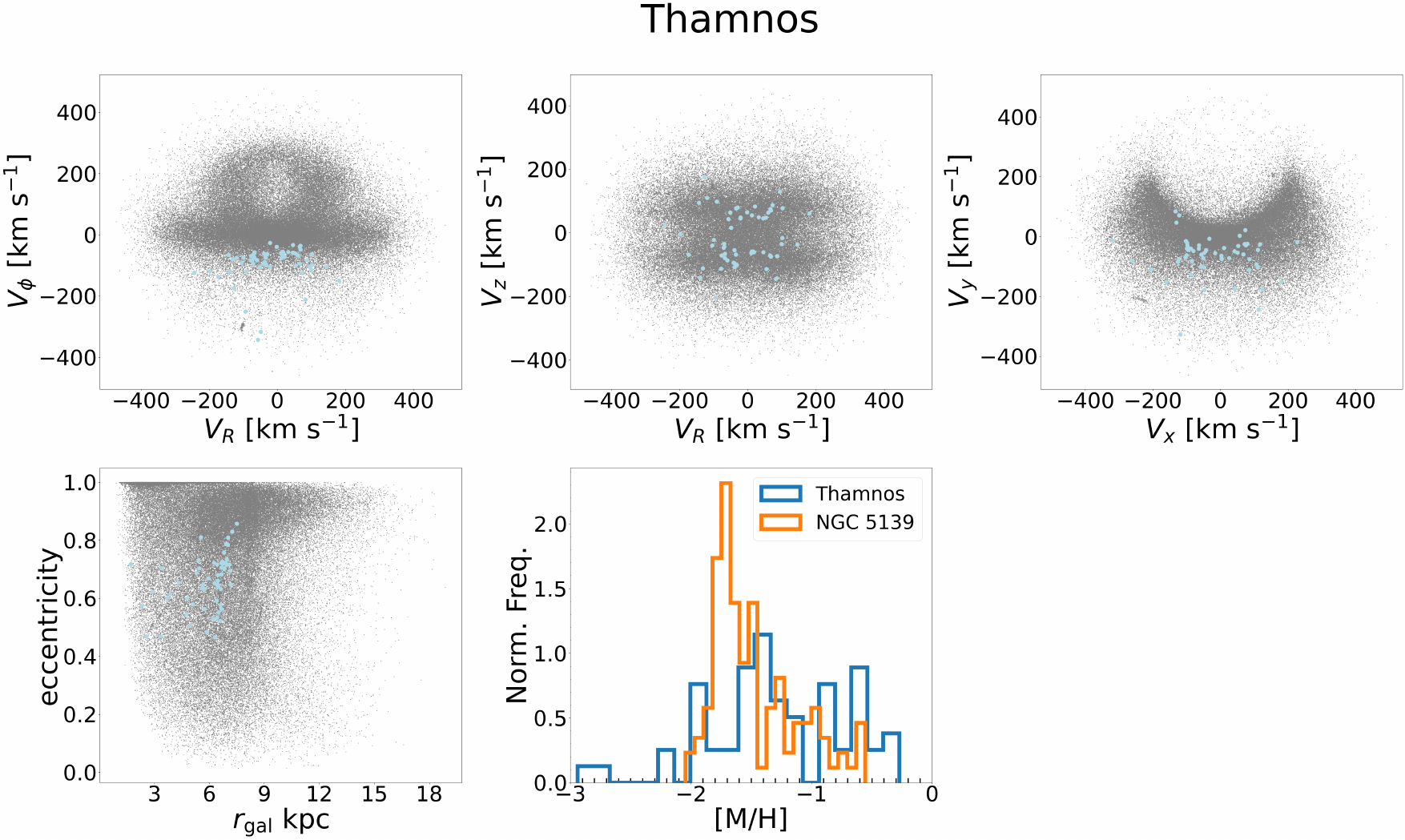} 
    \caption{Top row: the velocity distributions in $V_{R}-V_{\phi}$, $V_{R}-V_{x}$, $V_{x}-V_{y}$ space. Most Thamnos members have slight $|V_{\phi}|$, and have both positive and negative $V_{z}$. Bottom row: The eccentricity is ranging braodly from $\sim$ 0.35 to $\sim$ 0.85, with a visible metal-poor characteristic shown in MDF (the MDF of NGC 5139 is also shown for comparison). }
    \label{TH}
\end{figure*}

\begin{figure*}
    \includegraphics[width=\textwidth]{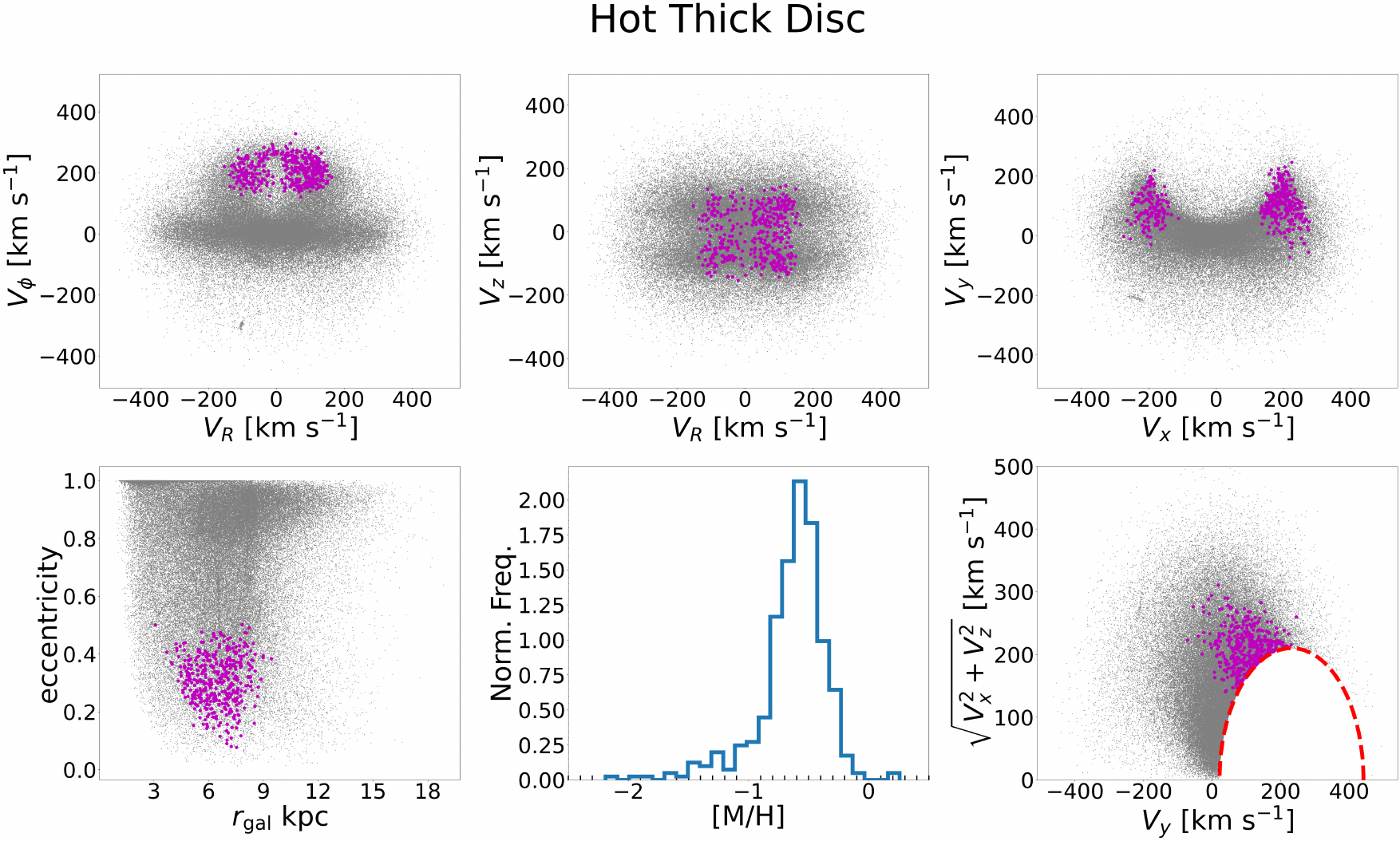} 
    \caption{ Top row:the velocity distributions in $V_{R}-V_{\phi}$, $V_{R}-V_{x}$, $V_{x}-V_{y}$ space. Bottom row: The eccentricity distributions along the galacocentric distance $r_{\text{gal}}$, metallicity distributions of Hot Thick Disc and Toomre diagram. The Hot Thick Disc is situated at the ends of banana-shaped arc in the $V_{X}-V_{y}$ plane. The close attachment to the curve where $|V-V_{\text{LSR}}|$ = 210 km/s is also visible in the bottom right panel. } 
    \label{HTD}
\end{figure*}

\begin{figure*}
    \includegraphics[width=\textwidth]{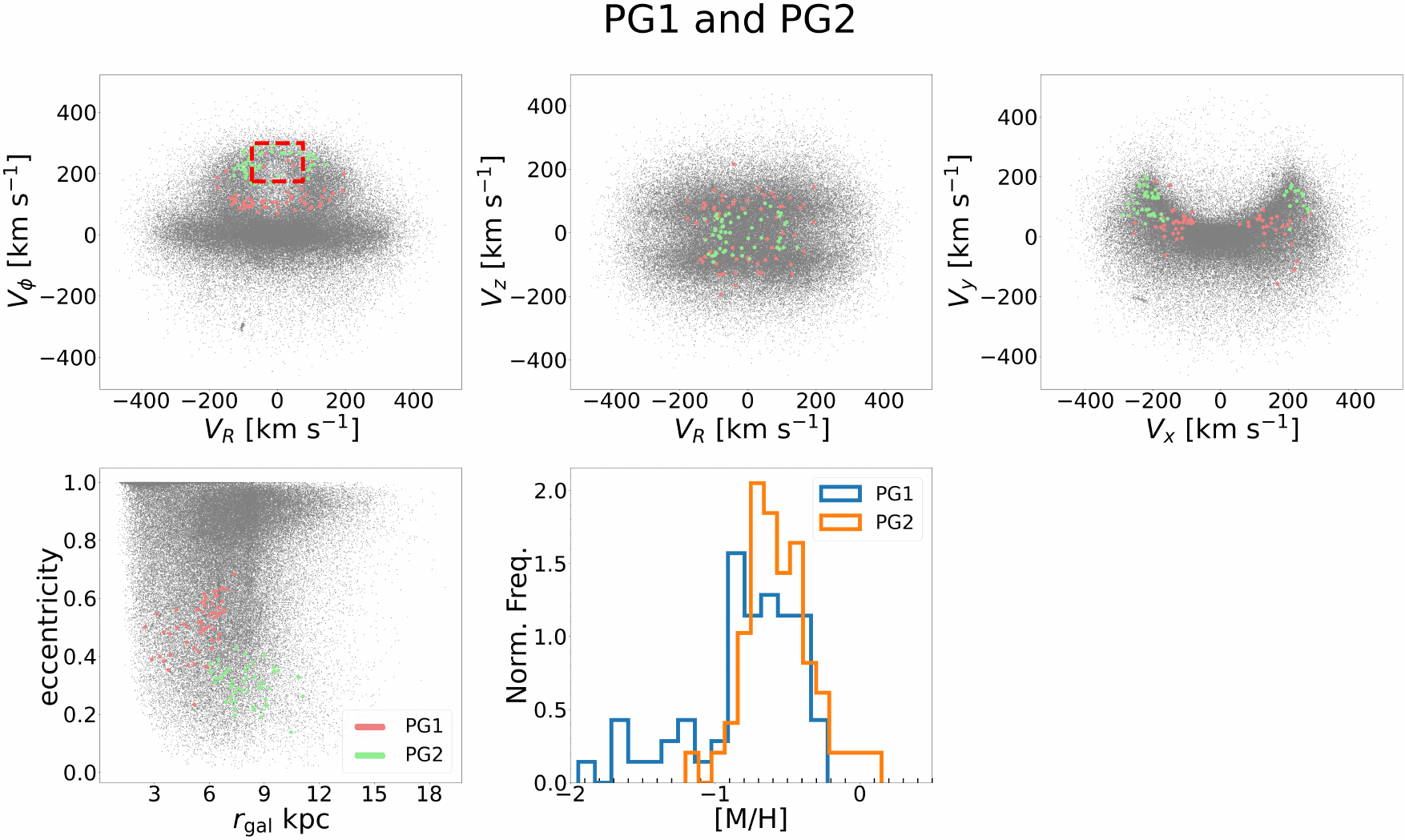} 
    \caption{Top row:the velocity distributions in $V_{R}-V_{\phi}$, $V_{R}-V_{x}$, $V_{x}-V_{y}$ space. Bottom row: The eccentricity distributions along the galacocentric distance $r_{\text{gal}}$ and metallicity distributions of PG1 and PG2. PG1 exhibits disc-like orbital properties, yet it features higher eccentricities, predominantly ranging from 0.4 to 0.6, along with a wide metallicity distribution. PG2 demonstrates Aleph-like kinematics and presents a natural transition within the eccentricity-$r_{\text{gal}}$ space, suggesting that it may represent the high eccentric and inner portion of the Aleph structure.} 
    \label{PG}
\end{figure*}

\begin{figure*}
    \includegraphics[width=\textwidth]{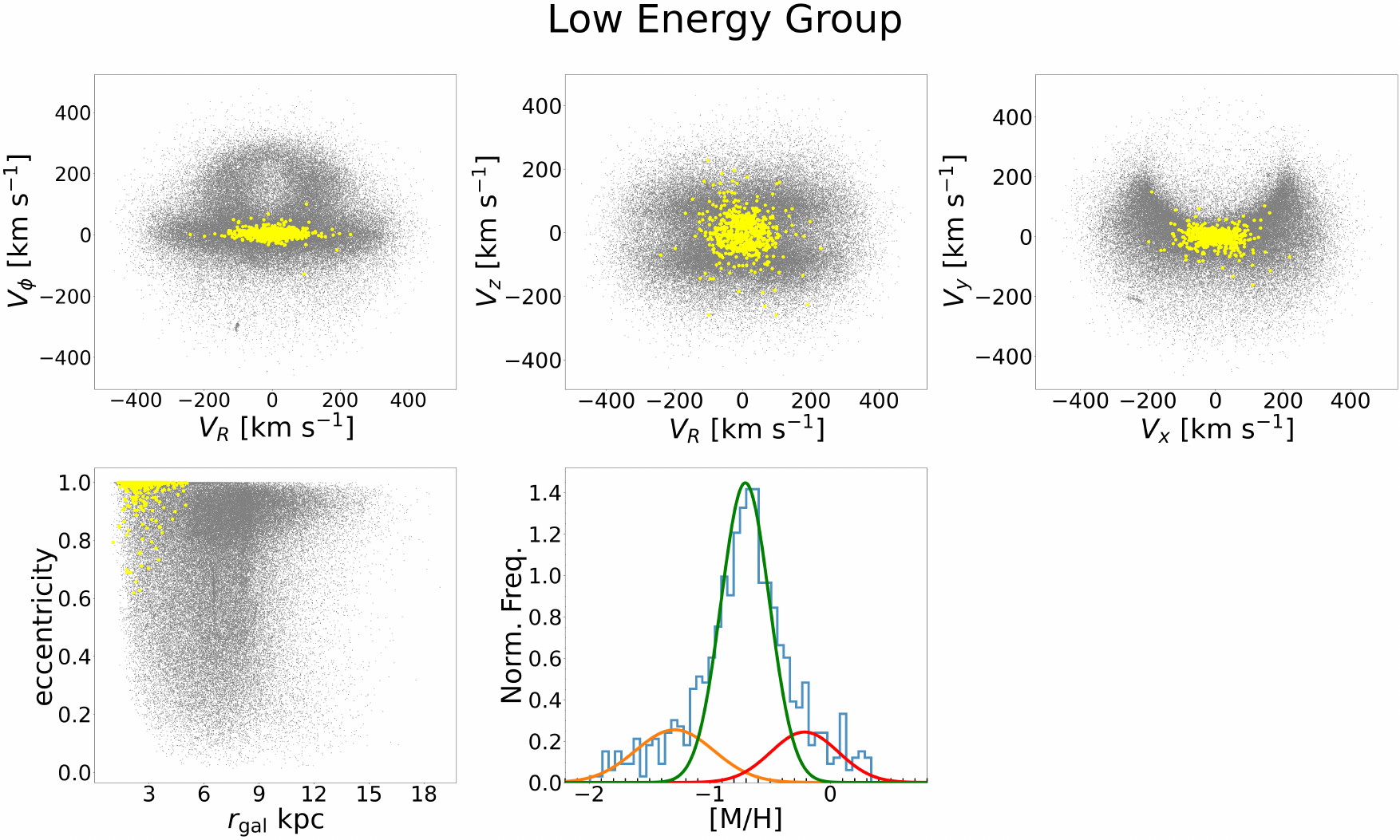} 
    \caption{Top row:the velocity distributions in $V_{R}-V_{\phi}$, $V_{R}-V_{x}$, $V_{x}-V_{y}$ space. Bottom row: The eccentricity distributions along the galacocentric distance $r_{\text{gal}}$ and metallicity distributions of Low Enegry Group. Low Energy Group have very small median values for $V_{R}$, $V_{\phi}$, $V_{x}$ and $V_{y}$, and they are $-$1.69 km/s, 4.76 km/s, $-$0.29 km/s and 1.47 km/s respectively. The MDF can be decomposed into three components, each with mean metallicity values [M/H] of approximately $-$1.29, $-$0.71, and $-$0.21, and corresponding weights of 19\%, 61\%, and 20\% respectively. The orange component with mean value $-$1.29 is in agreement with distribution of accreted region of Bulge in Horta's work, related to Heracles. } 
    \label{LEG}
\end{figure*}

\subsection{Helmi Streams}
Helmi Streams were discovered by \citet{HElim1999} who firstly applied the method of IoM, and was later researched again with updated \textit{Gaia} DR2 with an assumed progenitor mass of $10^{8}\ M_{\sun}$\citep{koppelmanB,koppelmanA,koppelman2021}. Helmi Streams stand out due to their high \textit{z}-velocities, which makes them easily identified in $L_{\perp}-L_{z}$ plane.

We identified a cluster with 94 stars as Helmi Streams and they are shown in Fig.~\ref{HS}. The prograde characteristic is illustrated in $E-L_{z}$ plane, with a median value of $V_{\phi}\sim$ 146 km/s, in good agreement with previous literature with a rough $V_{\phi}$ $\sim$ 150 km/s. Helmi Streams in our sample has $L_{\perp}$ $>$ 2000 kpc km s$^{-1}$ which makes it well-separated from other substructures (shown in Figure~\ref{ClusterR}). We also notice that Helmi Streams have negative $V_{z}$ velocities, which is thought as the second part, while the first part of Helmi Streams has positive $V_{z}$ velocities \citep{koppelmanB}. The first part is almost missing (only a few stars have positive $V_{z}$ velocities) because it is less populated and thus may not be well-detected by HDBSCAN. Helmi Streams exhibit a broad range of metallicity at the metal-poor end, with a median value [M/H] $\sim$ $-$ 1.54.

\subsection{The High Energy Retrograde Substructure (HERS): Arjuna (NGC 3201), I'itoi and Sequoia}
Arjuna and I’itoi were first identified as high-energy retrograde substructures from H3 survey by \citet{Naidu_2020}. Arjuna exhibits a mean [Fe/H] of $-$1.2, while I'itoi's [Fe/H] $<$ $-$ 2.0. I'itoi is more enriched in $\alpha$-elements compared to Arjuna and, conversely, has a lower $|L_{z}|$ than Arjuna. Sequoia is also a high energy retrograde substructure with [Fe/H] $\sim-$ 1.6, which may be the debris of a dwarf galaxy \citep{myeong_2019}.  However, recent simulations also show that Sequoia-like substructure can be generated by a single GSE-like merger event \citep{Naidu_2021,Amarante_2022}.

We identified initially one single cluster composed of 60 members as Arjuna, characterized by its extremely retrograde orbits in the $E-L_{z}$ plane, and another cluster with 50 members as the combination of I'itoi and Sequoia with less retrograde orbits. The identification is further confirmed by the MDF as shown in Fig.~\ref{HERS}.

Interestingly, Arjuna members almost occupy the same position both in phase and velocity space, with identical galactocentric distances in the $r_{\text{gal}}$ plane, which reminds us of Globular Clusters (GCs). We later confirmed that most of the cluster members indeed belonged to a globular cluster NGC 3201 , which was also discovered in action space by \citet{Ouxiaowei} and the name Arjuna will be replaced by NGC 3201 hereinafter. The median [M/H] value for NGC 3201 is $-$1.38, which is lower than Arjuna's [Fe/H] $\sim-$1.20. In recent research, NGC 3201 is considered as ex situ in \citet{Belokurov2024}. \citet{Munoz13} also studied 29 elements of NGC 3201, and this study shows that NGC 3201 has mean $\alpha$-abundances with [Mg/Fe] $\sim$ 0.38, [Ca/Fe] $\sim$ 0.32, [O/Fe] $\sim$ 0.13 and [Si/Fe] $\sim$ 0.25. The four $\alpha$-elements have a mean value $\sim$ 0.27, which is close to that of Arjuna [$\alpha$/Fe] $\sim$ 0.24. Combined with the almost same occupation in phase space and close $\alpha$ abundance value as Arjuna, they perhaps share the same origin.

As for I'itoi and Sequoia, they are clearly separated into two groups in the $V_{R}-V_{\phi}$ and $V_{R}-V_{z}$ plane: with positive $V_{R}$ and negative $V_{R}$, and does not show a continued span of $V_{R}$ as those in \citet{Naidu_2020}. We suggest two possible causes for this distribution: one is simply lack of samples (fewer than the samples in Naidu's work), and the other is due to its intrinsic instability (the substructure vanishes first as \texttt{min\_samples} increases). In their MDF, two distinct peaks are evident in the histogram: the first peak at [M/H] $\sim$ $-$ 2.00 and the second at [M/H] $\sim$ $-$1.6. We accordingly assign the first peak to I'itoi and the second to Sequoia. What needs stressing here is that in Naidu's work, the authors applied [Fe/H] cuts to define I'itoi ([Fe/H] $<$ $-$ 2.0), Sequoia ($-$ 2.0 $<$ [Fe/H] $<$ $-$ 1.5) and Arjuna ([Fe/H] $>$ $-$ 1.5) according to the clumps in MDF. However, these cuts could have cross-contamination in the samples as substructures cannot be completely restricted in the [Fe/H] bins, and so does the analysis of positions in IoM space. Consequently, there could be possible cross-contamination in our identified HERS.

\subsection{Thamnos}
Thamnos, a slightly retrograde substructure, constitutes a distinct clump positioned in the lower left quadrant of the $E-L_{z}$ distribution occupied by GSE. \citet{koppelman_2019} suggested that their progenitor was likely a dwarf galaxy accreted on a low-inclination orbit, and consistent with their low metallicity. Thamnos, which could be divided into two groups on the basis of metallicity and azimuthal velocity, named Thamnos 1 and Thamnos 2. 

There are two main criteria for selecting Thamnos members (with $E$ and $L_{z}$ already limited): one is the limitation on [Fe/H] $<$ $-$1.6 \citep{Naidu_2020}, which helps to exclude GSE stars with [Fe/H] peaking around $-$1.20, and the other is limitation on $e$ $<$ 0.7 \citep{Horta2023}. Given that many GSE member stars with high eccentricities have been identified and considering the potential systematic errors in metallicities, we take the second approach and identify a slightly retrograde cluster as Thamnos. However, we noticed the overdensity in the eccentricity$-r_{\text{gal}}$ distribution again and identified some stars belonging to NGC 5139, a globular cluster that is likely an in situ Nuclear Star Cluster but with chemical abundance ratios similar to accreted systems \citep{Belokurov2024}. We excluded 116 NGC 5139 member stars with 59 stars left in the Thamnos samples. 

Fig.~\ref{TH} shows the distributions of Thamnos after removing NGC 5139. The majority of Thamnos stars exhibit $V_{\phi} < -$26 km/s, indicating a slight retrograde motion. The MDF has a median value of $-$1.36 with clearly metal-poor characteristics.
Thamnos also shows a wide span of eccentricities $\sim$ [0.4, 0.9], that are averagely higher than Thamnos samples $\sim$ [0.2, 0.7] in \citet{Naidu_2020}.

\subsection{Hot Thick Disc}
Hot Thick Disc describes stars that are kinematically hotter than normal disc stars, which can contaminate the halo samples with only Toomre cut. The Hot Thick Disc is vividly characterized as banana-shaped in $V_{x}-V_{y}$ plane and is closely attached to the curve where $|V-V_{\text{LSR}}|$ = 210 km/s as shown in Fig.~\ref{HTD} \citep{Koppelman_2018,Helmi_2020}. We identified three clusters with 412 stars as the Hot Thick Disc according to its unique characteristics as well as disc-like behaviors.

The MDF exhibits a single peak and a long metal-poor tail towards [M/H] $\sim -$ 0.2. Hot Thick Disc has a median [M/H] value of approximately $-$0.60, similar to the [Fe/H] value of about $-$0.64, as indicated by APOGEE data in \cite{DOdd_2022}. The median rotational velocity $V_{\phi}$ of our Hot Thick Disc is approximately 205 km/s, which is the characteristic of typical disc rotation speeds \citep{Yan2020,Zhu2021}. The radial velocity $V_{R}$ in our samples exhibits a relatively continuous distribution, ranging from $\sim -$150 km/s to $\sim$ 160 km/s, with a median value of $\sim$ 53 km/s. This is distinct from the result reported by \citet{DOdd_2022}, where the Hot Thick Disc is situated around the two ends of the $V_{R}$ at approximately $-$ 200 km/s and 200 km/s, respectively. More detailed and illustrative information can be found in Fig.~\ref{HTD}.

\subsection{Prograde Substructure 1 and Prograde Substructure 2 }
In addition to the previous explored substructures, we have also discovered three new substructures in this study. Two of these exhibit prograde characteristics and they are named Prograde Substructure 1 (with a lower $|L_{z}|$, single cluster with 61 stars) and Prograde Substructure 2 (with a higher $|L_{z}|$, single cluster with 54 stars), hereinafter referred to as PG1 and PG2, respectively. The information for PG1 and PG2 is shown in Figure~\ref{PG}, and we provide the relevant analysis below.

PG1 shows a slight prograde property with a median value of $L_{z}\sim$ 579 kpc km s$^{-1}$, and a median value of $V_{\phi}$ $\sim$ 111 km/s,  which is significantly higher than that of the GSE.  PG1 shows a disc-like distribution in the $V_{R}-V_{\phi}$ plane, but with higher eccentricities, mostly ranging from 0.4 to 0.6, which exceeds the range of 0.2 to 0.4 typical of the Hot Thick Disc. Considering this, PG1 might belong to the thick disc but with higher radial velocities and lower tangential velocities, which distinguishes them from the Hot Thick Disc. However, according to the metallicity distribution, PG1 exhibits a broad range in metallicity, spanning from [M/H] $\sim$ $-$1.95 to $-$0.22, with the majority concentrated between $-$ 1.0 and $-$ 0.6, and showing no distinct peak. Simulation shows that for a GSE-like merger, accreted stars with low eccentricities tend to be metal-rich, referring to Figure 6 in \citet{Amarante_2022}. Considering the proximity of PG1 to GSE in phase space, it is plausible that PG1 represents the prograde and metal-rich component of GSE. 

PG2 shows a far more prograde property with $L_{z}\sim$ 1800 kpc km s$^{-1}$, and shares kinematic similarities with Aleph as shown in the $V_{R}-V_{\phi}$ plane, where the dashed red lines indicate the selection criterion that \citet{Naidu_2020} uses $|V_{R}| <$ 75 km/s, 175 km/s $<$ $V_{\phi}$ $<$ 300 km/s to select Aleph. Most of the PG2 member stars are situated close to the lines, with a slightly larger span in $V_{R}$. In the eccentricity-$r_{\text{gal}}$ plane, PG2 is distributed with eccentricities ranging from $e \sim 0.1$ to $e \sim 0.4$ and galactocentric distances from $r_{\text{gal}} \sim 6$ kpc to $r_{\text{gal}} \sim 11$ kpc. In contrast, most Aleph stars spread out from 7 kpc to 17 kpc, exhibiting eccentricities lower than 0.3. Approximately 91\% of PG2 stars exhibit metallicities greater than $-$ 0.8, which almost meets the criterion for Aleph stars, which necessitates metallicities of no less than $-$ 0.8. The similarities in kinematics, metallicities and the smooth transition observed in the eccentricity-$r_{\text{gal}}$ plane, indicate that PG2 might be more eccentric and spatially inner component of Aleph. It could also be the mixture of Aleph and disc due to its low $Z_{\text{gal}}$ (with a median value of 1.23 kpc compared to 3.51 kpc of Aleph). This suggestion calls for additional confirmation through the analysis of chemical abundances.

\subsection{The Low Energy Group}

As stated in subsection~\ref{G-S-E}, we excluded six clusters containing 641 stars from the initial GSE sample and identified them as the Low Energy Group, due to their extremely eccentric orbits and, of course, their low energy. The galactocentric distances of this substructure range from 1 kpc to 5 kpc, which is clearly in the Galactic bulge. \citet{Horta2021} identified Heracles in the [Mg/Mn]-[Al/Fe] plane using APOGEE sample within 4 kpc of galactocentric distances. Heracles exhibits a distinct energy gap compared to the GSE and is likely associated with the low-energy globular clusters accreted from a dwarf galaxy known as Kraken (or Koala) \citep{K19,K20,Forbes_2020,Naidu_2022}.

The kinematic properties of Low Energy Group is shown in Fig.~\ref{LEG}. The Low Energy Group member stars show homogeneous distributions in the velocity space (i.e. $V_{R}-V_{\phi}$, $V_{R}-V_{z}$ and $V_{x}-V_{y}$ planes) and high eccentricities, which are condensed in the area near the bulge, as noticeable in the eccentricity-$r_{gal}$ plane. The MDF of Low Energy Group is much more complex, filled with multiple stellar populations of both metal-poor and metal-rich stars. Therefore, we applied the Gaussian Mixture Model from the Python scikit-learn package \citep{scikit-learn} to decompose the metallicity distribution. We determined the optimal number of components to be three based on the Bayesian Information Criterion (BIC). This result is similar to the bulge MDF described in \citet{Horta2021}, where they found that the accreted component constitutes the metal-poor fraction of the bulge MDF, alongside the in-situ high-$\alpha$ and low-$\alpha$ components. This similarity is not surprising, as the majority of the Low Energy Group stars are situated within the bulge. The orange line (bottom middle panel of Fig.~\ref{LEG}) indicates the metal-poor component with a median value of [M/H] $\sim$ $-$1.29 similar to the mean value $-$1.26 of Heracles. We therefore think that Low Energy Group may include part of Heracles, and the reason why HDBSCAN did not detect Heracles was probably due to its small weight ($\sim$ 19\%) in the samples.

Although the MDF of the Low Energy Group exhibits a wide spread, the member stars display surprisingly similar kinematics and high eccentricities, thus as  overdensities identified by HDBSCAN. However, whether these stars are in-situ within the bulge, or ex-situ, originating from an accretion event and represent as a substructure, remains to be confirmed by future surveys.

\section{Discussion and Summary}\label{discussion}
Based on the precise parallaxes, proper motions and radial velocities from \textit{Gaia} DR3 in combination with the well-estimated metallicities from \textit{Gaia} XP, we investigated halo substructures using the IoM space, which is defined by the parameters \(E\), \(L_{z}\), and \(L_{\perp} = \sqrt{L_{x}^2 + L_{y}^2}\). We used Hierarchical Density-Based Spatial Clustering of Applications with Noise for its efficiency and fewer hyper-parameters, especially in situations with an unknown number of clusters and nonuniform data density. After parameter adjustment, 30 groups containing 2,932 stars associated with 9 substructures were obtained.

In this study, we omit chemical elements abundances from high-resolution surveys due to limited coverage and insufficient cross-matched data for statistical significance.  Although there may be doubts due to the lack of chemical information, our methodology of using the IoM is effective means to identify substructures. The conservation of these quantities, with $L_{\perp}$ evolving slowly over time, preserves the orbital "memories" of substructures inherited from their parent galaxies. Despite such concerns, the application of IoM space, alongside the precise metallicity estimates from Gaia XP spectra achieved by XGBoost, and good matches in both phase and velocity space in previous literature, proves to be a robust and valid tool.

However, we failed to identify substructures like Sagittarius stream, LMS-1 (Wukong) and  Pontus.  The Sagittarius stream is barely visible within our samples, which are confined to $<$ 19 kpc, as it is most prominent beyond $\sim$30 kpc. Its inner Milky Way debris is too sparse to exhibit a clear overdensity in phase space and therefore we found no corresponding substructure in this work. As for LMS-1 (Wukong) and Pontus, they overlap with GSE and Metal Weak Thick Disc (MWTD) in phase space, sharing obvious metal-poor characteristics \citep{Ye2024}. Without metallicities and chemical abundances considered in the clustering analysis, it is hard to differentiate substructures occupying the same IoM space, and that is why substructures requiring multi-dimensional analysis beyond IoM were not identified.

In summary, we successfully used HDBSCAN method to identify five previously discovered substructures, two Globular Clusters (NGC 3201 and NGC 5139) and three unknown substructures. The five existing substructures are Gaia-Sausage-Enceladus (GSE), Helmi Streams, Thamnos, Sequoia and I'itoi (HERS, Sequoia may have a different accreted origin from GSE) and Hot Thick Disc. NGC 3201 shows similar distributions to Arjuna's and likely has the same accreted origin, so it is interchangeably referred to as Arjuna in this context, while NGC 5139 is removed from initial Thamnos samples for its possible in situ origin. The metallicity distribution function of GSE displays two peaks, with the first peak at [M/H] $\sim-$ 1.25 and the second peak at [M/H] $\sim-$ 0.7, suggesting contamination from Splash stars with [M/H] $>$ $-$ 0.7. The slight retrograde Thamnos lies near the central region where GSE is located in the $E-L_{z}$ plane. Helmi Streams are the clearest substructure, standing out in the phase space with its prominent $L_{\perp}$. Arjuna (NGC 3201), I'itoi and Sequoia complement the retrograde halo structures, showing high energy and metal-poor characteristics. 

These new substructures identified from IoM space are: Prograde Substructure 1 (PG1), Prograde Substrcuture 2 (PG2) and Low Energy Group. PG1 shows disc-like orbits while showing higher eccentricities, may belong to metal-poor part of the thick disc with high radial velocities, or belong to the less eccentric and metal-rich part of GSE. PG2 is more prograde than PG1 and exhibits Aleph-like characteristics in velocity space, showing a smooth transition in eccentricities and galactocentric distances and it is considered as the spatially inner part of Aleph or the mixture of Aleph and disc . The Low Energy Group with eccentricities approaching \textit{e} $\sim$1, is speculated to have part of Heracles, representing a possible new substructure. These new substructures require further confirmation with the advent of future surveys.

\section{ACKNOWLEDGENMENTS}
\par This work was supported by the National Natural Science Foundation of China (NSFC Nos: 11973042, 11973052). This work has made use of data from the European Space Agency (ESA) mission
{\it Gaia} (\url{https://www.cosmos.esa.int/gaia}), processed by the {\it Gaia}
Data Processing and Analysis Consortium (DPAC,
\url{https://www.cosmos.esa.int/web/gaia/dpac/consortium}). Funding for the DPAC
has been provided by national institutions, in particular the institutions
participating in the {\it Gaia} Multilateral Agreement.

\bibliography{lhy}{}
\bibliographystyle{aasjournal}



\end{document}